\documentclass[aps,prb,twocolumn,a4paper,showkeys,showpacs, superscriptaddress]{revtex4-1}

\usepackage{graphicx}
\usepackage{mediabb}
\usepackage{comment}
\usepackage{color}
\usepackage{textcomp}

\begin{document}

\title{First-principles calculations that clarify energetics and reactions of oxygen adsorption and carbon desorption on 4H-SiC ($11\bar20$) surface}
\author{Han \surname{Li}}
\affiliation{Department of Applied Physics, The University of Tokyo, Hongo, Tokyo, 113-8656, Japan}
\author{Yu-ichiro \surname{Matsushita}}
\affiliation{Department of Applied Physics, The University of Tokyo, Hongo, Tokyo, 113-8656, Japan}
\author{Mauro \surname{Boero}}
\affiliation{Institut de Physique et Chimie des Mat\'eriaux de Strasbourg, University of Strasbourg-CNRS UMR 7504, 
23 rue du Loess, F-67034 Strasbourg, France}
\author{Atsushi \surname{Oshiyama}}
\affiliation{Department of Applied Physics, The University of Tokyo, Hongo, Tokyo, 113-8656, Japan}

\begin{abstract}

We report static and dynamic first-principles calculations that provide atomistic pictures 
of the initial stage of the oxidation processes occurring at the ($11\bar20$) surface of 4H-SiC. 
Our results unveil reaction pathways and their associated free-energy barriers for the adsorption 
of oxygen and the desorption of carbon atoms. 
We find that oxygen adsorption shows structural multi-stability and that the surface-bridge 
sites are the most stable and crucial sites for subsequent oxidation. 
We find that an approaching O$_2$ molecule is adsorbed, then dissociated and finally migrates 
toward these surface-bridge sites with a free-energy barrier of 0.7 eV at the ($11\bar20$) surface.
We also find that a CO molecule is desorbed from the metastable oxidized structure 
upon the overcoming of a free-energy barrier of 2.4$\sim$2.6 eV, 
thus constituting one of the annihilation process of C during the oxidation. 
The results of the CO molecule desorption on the ($11\bar20$) surface are compared with the ($000\bar1$) surface.
A catalytic effect of dangling bonds at the surface, causing a drastic reduction of the CO desorption energy, 
is found on the ($000\bar1$) surface 
and the microscopic picture of the effect is ascribed to an electron transfer from the Si to C dangling bonds. 
The intrinsic ($11\bar20$) surface does not show this catalytic effect, 
and this is because the surface consists of an equal amount of Si and C dangling bonds 
and the electron transfer occurs before the desorption.

\end{abstract}

\maketitle

\section{Introduction}

Oxidation of materials is a fundamental phenomenon in nature. In particular, the oxidation of semiconductor surfaces 
usually leads to the formation of insulating films on the semiconductors. 
The resultant interface of the oxide and the semiconductor is essential in the transistor action of semiconductor devices. 
For instance, silicon dioxide (SiO$_2$) formed on Si is a good insulator, thus sustaining current technology.

Silicon carbide (SiC) is a well-known and now an emerging semiconductor material for power electronics due to 
its wide band-gap and the robustness under harsh environment,\cite{SiC-Strain-Sensor, Soo201039} possibly succeeding the premier Si in the future. 
Insulating films for the SiC device are also SiO$_2$ formed by the thermal oxidation. 
However, the formation of the SiO$_2$ films on the compound semiconductor SiC is a much more complicated phenomenon 
since oxygen reacts with both Si and C, and then C atoms are eventually annihilated. 
Clarification of the microscopic mechanism for the oxidation of SiC is therefore extremely interesting.

SiC devices such as schottky diodes and field-effect transistors are already fabricated and in production.\cite{Yu2009-Schottky, 6064129-MOSFET} 
However, there still remains much room for the improvement when considering the good bulk properties of SiC: 
e.g., the electron mobility in the devices is typically less than 10\% of the bulk value.\cite{Interface_mobility, SiC-review} 
This obviously comes from the poor controllability of the SiO$_2$/SiC interface formed by the oxidation. 
In order to achieve the microscopic identification and then increase the controllability of the interface, 
intensive works have been done both experimentally\cite{SiC-exp1, SiC-exp2, SiC-exp3, SiC-exp4, SiC-exp5, SiC-exp6, SiC-exp7, Interface_mobility} 
and theoretically\cite{SiC-theory1, SiC-theory2, SiC-theory3, SiC-theory4, SiC-theory5, SiC-theory6, SiC-theory7}. 
However, there is still no definite explanation for the deterioration mechanism of electron mobility. 
It is thus highly demanded to perform accurate calculations which clarifies the mechanism of 
the oxidation of SiC surfaces and then provides a theoretical framework to discuss the nature of the SiO$_2$/SiC interface.

The purpose of the present paper is to focus on the initial stage of the oxidation of SiC 
and provide its atomistic reaction mechanism based on the first-principles calculations. 
The initial stage already includes salient features of the oxidation such as the selective oxygen attack of 
either C or Si atom and the C annihilation by plausible reaction pathways. 
SiC is a covalent and at the same time polar semiconductor and exists as various polytypes. 
The most stable and the commonly used is 4H-SiC in which the atomic layers perpendicular to 
the bond direction stack with the quad periodicity causing the hexagonal symmetry (4H). 
We focus on the ($11\bar20$) surface of 4H-SiC in this paper since the devices 
with high electron mobility are fabricated on the surface.\cite{Device_mobility} 
Also the ($11\bar20$) surface is non-polar surface on which both Si and C atoms appear, 
being a good stage to reveal the affinity of O with Si and C during the oxidation process.

The initial stage of the oxidation process is decomposed into three main steps: (i) the O$_2$ molecular adsorption 
and subsequent dissociation on the surface, (ii) the migration of the atomic O toward the formation of the Si-O 
bond network, and (iii) the C removal from the surface as either a monoxide CO or a dioxide CO$_2$. We clarify 
the microscopic mechanism of these processes by the static total-energy calculations and via dynamic first 
principles molecular dynamics simulations within the density-functional theory (DFT)\cite{Hohenberg-Kohn, Kohn-Sham} framework 
according to the Car-Parrinello scheme (CPMD).\cite{CPMD}
To explore the reaction pathways and obtain the corresponding free-energy barriers, we complement CPMD with 
the meta-dynamics\cite{MTD} approach.

The organization of this paper is as follows. 
Section II shows pertinent features of the present static and dynamic calculations. 
Results and discussion are presented in sections III, IV and V. 
We first present various stable and metastable atomic configurations for the oxygen adsorption obtained by the static calculations in III A and III B. 
Then the dynamical aspects of the O$_2$ molecular adsorption, dissociation and the O migration are shown in IV A and IV B. 
The reaction of the carbon oxide desorption from the surface is investigated in V A and V B. Section VI summarizes our findings.

\section{Methodology and slab model}

\subsection{Static Calculations}

Stable atomic configurations for the oxygen adsorption on the 4H-SiC ($11\bar20$) surface have been explored by the total-energy electronic-structure calculations 
based on DFT, with generalized gradient approximation by Perdew, Burke and Ernzerhof (PBE)\cite{PBE96} 
for the exchange-correlation energy functional. 
Nuclei and core electrons are treated by the projector-augmated wave (PAW) scheme by Bl\"ochl.\cite{PAW} 
The plane-wave basis set is used to express the Kohn-Sham orbitals and thus the electron density and the self-consistent potentials. 
The convergence of the basis set has been examined with bulk Si and C, and molecular O. 
The cutoff energy of 400 eV is found to be enough to assure the accuracy in the total-energy difference of 2 meV per atom. 
We simulate the ($11\bar20$) surface by the repeating slab model in which 6 atomic layers constitute the slab 
and each slab is separated from its adjacent images by the 15 \AA -thick vacuum region. 
We use the 1$\times$2 lateral cell which is a square of the dimension of about 10 \AA\ (see below). 
Correspondingly, the Brillouin zone (BZ) integration is performed by using the $3\times3\times1$ Monkhorst-Pack k-point sampling,\cite{MP-kpoint} 
which ensures the total-energy difference of 5 meV per cell. 
Computations have been done using the Vienna Ab-initio Simulation Package (VASP).\cite{VASP1, VASP2, VASP3, VASP4}

\subsection{Dynamic Calculations}

Dynamical aspects and thermal effects of the O$_2$ molecular adsorption, the subsequent O atomic migration and 
eventually the CO molecular desorption have been accounted for by CPMD simulations complemented with the 
meta-dynamics approach for the sampling of the free energy landscape.\cite{CPMD-code1, CPMD-code2, CPMD-code3} 
Norm-conserving pseudopotentials\cite{NCPP} are adopted to treat the core--valence interactions for H, C, O and Si. 
The PBE functional is used for the exchange-correlation interaction. The Kohn-Sham orbitals are expanded 
in a plane-wave basis set with a cutoff energy of 80 Ry ($\approx$1000 eV). 
In all canonical simulations, the ionic temperature is controlled with Nos\'e-Hoover thermostat.\cite{Nose, Hoover} 
A simulation time step of 4 a.u. ($\approx$0.1 fs) and a fictitious electronic mass for the wave functions of 300 a.u. 
ensure good control of the conserved quantities along the dynamics on the ps time scale used here.

To explore the reaction pathways within the metadynamics approach, suitable reaction coordinates (collective 
variables, CVs) being able to account for all the slowly varying degrees of freedom have to be selected.
To this aim, we choose as CVs the distance between the two O atoms of the adsorbed O$_2$ molecule and 
the distance between one O atom and the target sites for the O$_2$ molecular dissociation. 
Instead, for the CO molecular desorption, the coordination number\cite{COORD} of the C atom of the desorbing CO molecule 
is the selected CV. The expression of the coordination number is,
\begin{equation}
CN_i=\sum^{}_{j\neq i}\frac{1}{1+e^{k(d_{ij}-d^0)}}
\end{equation}
where $i$ is the index of the selected atom, $j$ runs over the surrounding atoms in the system, $d_{ij}$ is the distance 
between atom $i$ and atom $j$, $k$ determines the steepness of the decay of the analytical function shown above and $d^0$ 
is the reference distance. The index $j$ can be chosen to run over all the atoms of a specific 
chemical element. From the radial distribution function of SiC and diamond, we choose the reference distance of 
C-Si to be 2.2 \AA\ and C-C to be 1.8 \AA.

\subsection{4H-SiC Slab Models}

SiC exists as various polytypes. The structural difference lies only in the stacking of atomic planes along the bond direction, 
i.e., the axis of the inherent three-fold rotational symmetry: 
e.g., the stacking sequences of ABC in the zincblende and AB in the wurtzite structures, respectively. 
The stacking sequence is not limited to the above two cases. 
Hence there are dozens of polytypes labeled by the periodicity of the stacking sequence $n$ 
and the symmetry (cubic or hexagonal) such as 2H (wurtzite), 3C (zincblende), 4H and 6H. 
Among those, the 4H-SiC is the most stable and commonly used in the device applications. 
Each atomic site in the 4H structure is inequivalent to each other and classified into two groups: 
the site with the locally hexagonal symmetry (h-site hereafter) and the site with the locally cubic symmetry (k-site hereafter).

\begin{figure}
	\includegraphics[width=0.4\textwidth]{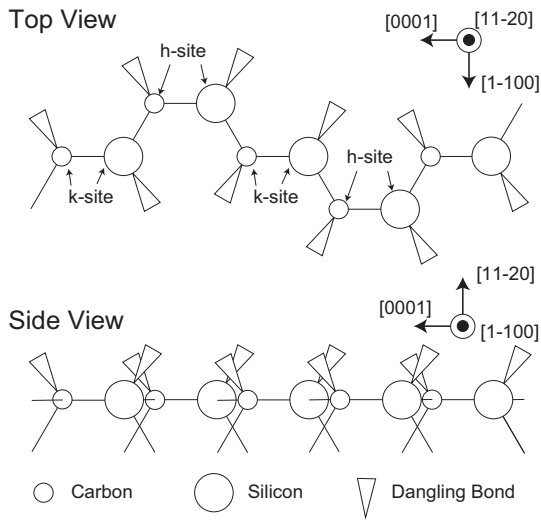}
	\caption{Schematic picture of the surface structure of 4H-SiC ($11\bar20$) with dangling bonds explicitly shown. The silicons and carbons are sp$^3$ hybridized thus the dangling bonds on the ($11\bar20$) surface are not perpendicular to the surface but slightly tilted. The constituent atoms of 4H-SiC are categorized into two kinds, h-site and k-site, as shown in the top view.}
	\label{fig_1120db}
\end{figure}

On the ($11\bar20$) surface, in addition to the appearance of both Si and C atoms, 
the h- and k- sites emerge on the topmost atomic plane alternately along [0001] direction. 
The schematic figure of ($11\bar20$) surface with dangling bond direction is shown in Fig. \ref{fig_1120db}.

\begin{figure}
	\includegraphics[width=0.5\textwidth]{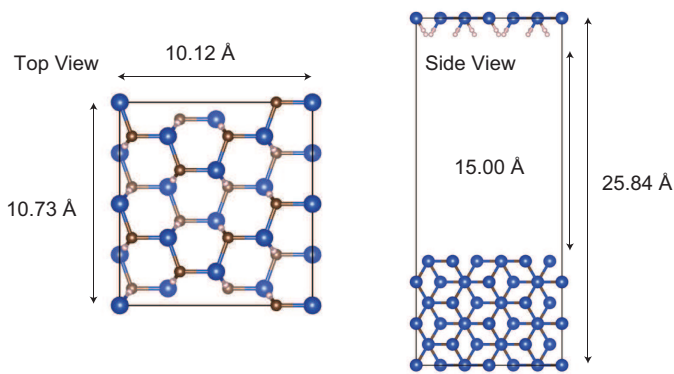}
	\caption{Top view and side view of ($11\bar20$) surface model. The bottom SiC layer and the hydrogen atoms are fixed during the geometry optimization and the CPMD simulation. The blue, brown and white balls depict Si, C and H atoms, respectively.}
	\label{fig_1120}
\end{figure}

Repeating slab models are used to simulate the SiC surface. 
Adjacent slabs are isolated by the vacuum region with the thickness of 15 \AA\  to avoid the fictitious interaction between the slabs. 
For the ($11\bar20$) surface, we have used $1\times2$ lateral unit cell and each slab contains 6 atomic layers with the thickness of 7.5 \AA. 
The bottom surface is terminated by H atoms to remove the fictitious dangling bonds electronically. 
The H atoms and the bottom  SiC layer are fixed during the geometrical optimization. The top view and side view of this model are shown in Fig. \ref{fig_1120}.

\begin{figure}
	\includegraphics[width=0.5\textwidth]{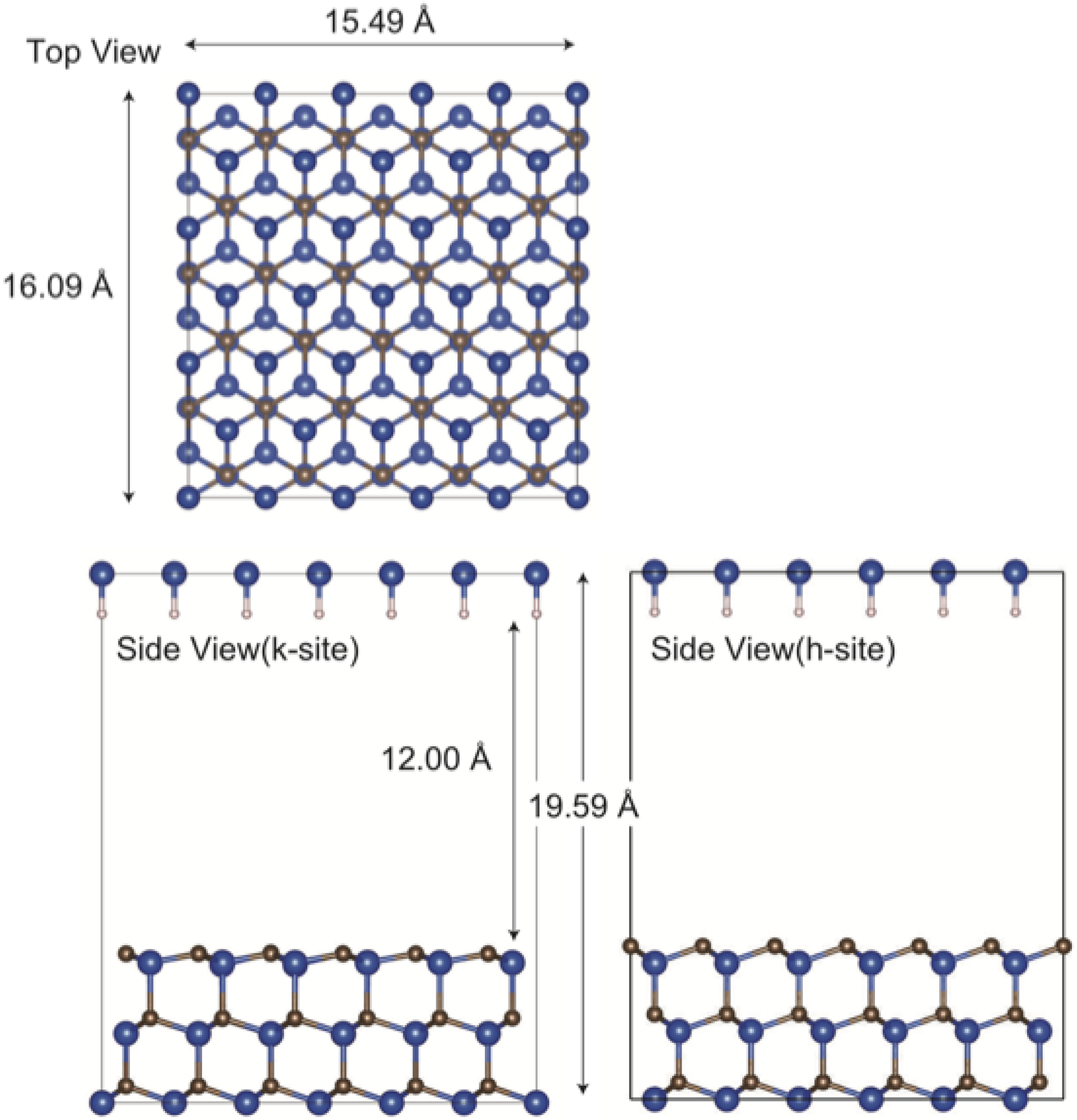}
	\caption{Top view and side view of ($000\bar1$) surface model. The bottom Si layer and hydrogen atoms are fixed during the geometry optimization and the CPMD simulation. The color code is the same as in Fig. \ref{fig_1120}.}
	\label{fig_0001}
\end{figure}

We also perform the calculation for the ($000\bar1$) surface to compare the characteristics of the initial oxidation. 
The adsorption sites of the O atom on ($000\bar1$) has already been studied.\cite{SiC_0001} 
For our purpose, we use the slab model in which each slab consists of 3 atomic bilayers, 
i.e., 6 atomic layers with the 5.5 \AA\ thickness, with $5\times6$ lateral periodicity. 
The vacuum thickness used is 12 \AA. The bottom surface is again terminated by H atoms. 
As there are two kinds of ($000\bar1$) surface, with h-site C and k-site C as its topmost atom plane, 
we perform the calculation for both. The top view and side view of the slab is shown in Fig. \ref{fig_0001}. 
The two kinds of ($000\bar1$) surfaces have the same top view but the difference is in the side view.

\section{Result I: adsorption of O atoms}

In this section, we present stable and metastable structures of the oxygen adsorption on the ($11\bar20$) surface of 4H-SiC obtained by the geometry optimization in DFT. We consider the one-O-atom and the two-O-atom adsorptions. We have found various stable structures which provide the energy landscape for the oxygen adsorption.

\subsection{Single oxygen-atom adsorption}

\begin{figure}
	\includegraphics[width=0.5\textwidth]{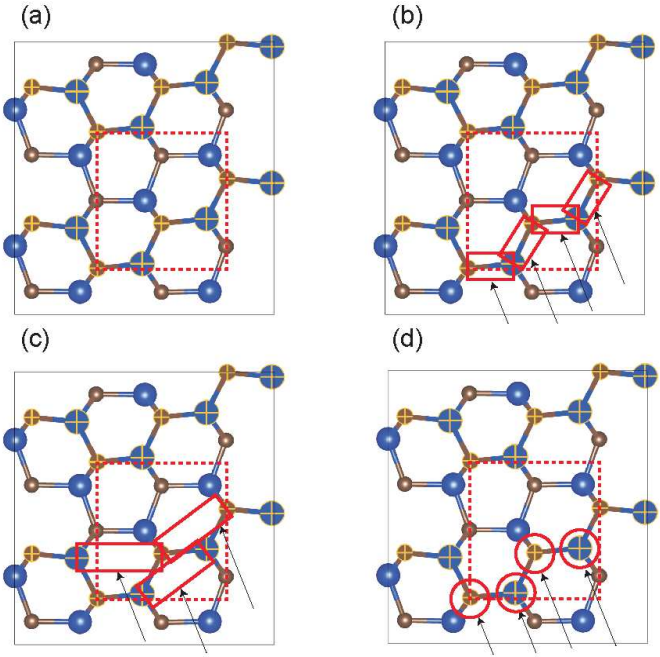}
	\caption{Initial structures in the exploration of the single O-atom adsorption sites. The color code is the same as in Fig. \ref{fig_1120}. The dotted lines depict the primitive lateral cell. The topmost atoms are marked by the crosses. (a) clean surface. (b) 4 bridge sites between the nearest-neighbor topmost atoms. (c) 3 bridge sites between the next neighbor or next-next neighbor topmost atoms. (d) 4 on-top sites. There are also 4 bridge sites between these topmost atoms and the subsurface atoms.}
	\label{fig_init-1O}
\end{figure}

\begin{figure}
	\includegraphics[width=0.5\textwidth]{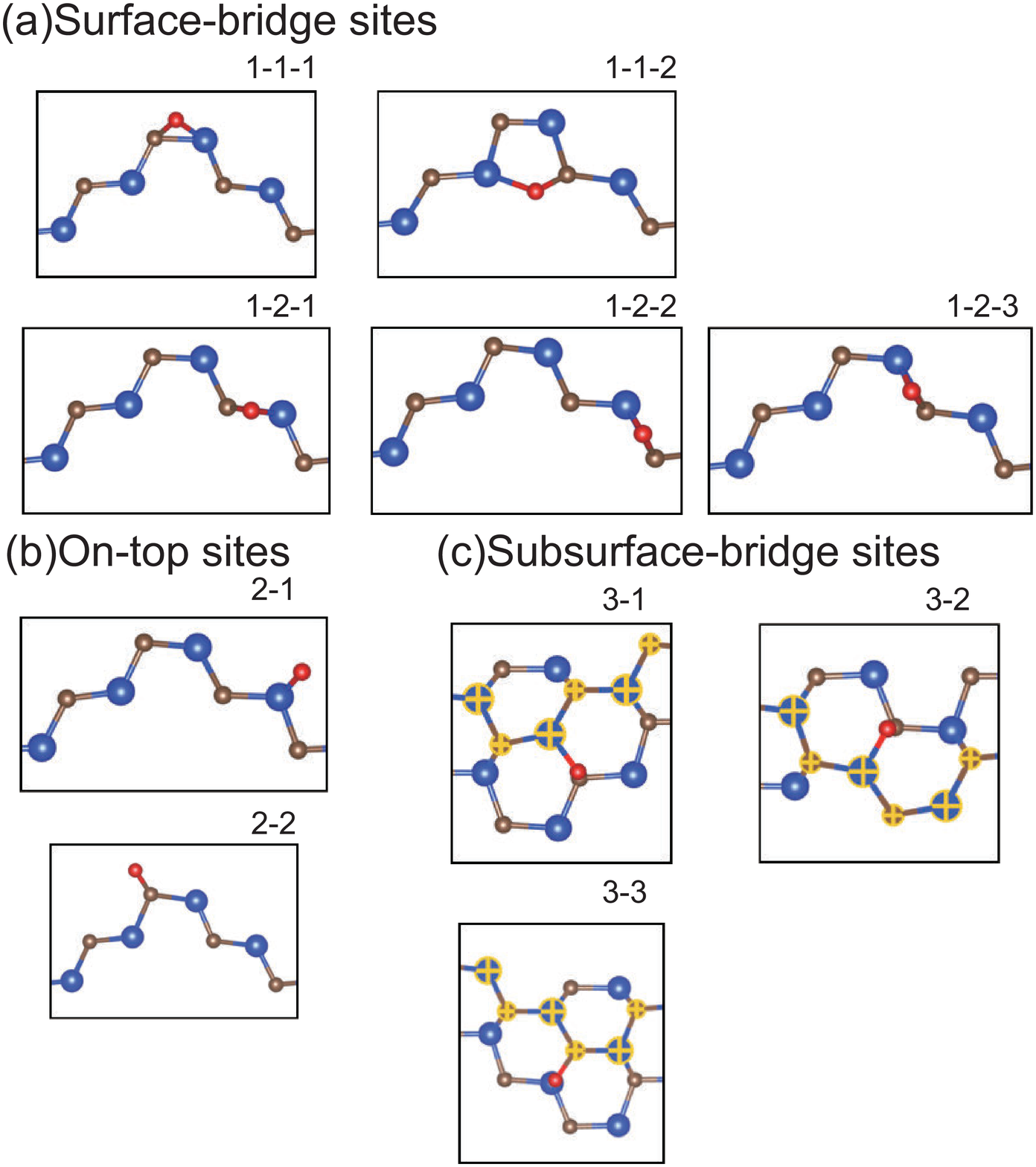}
	\caption{Top views of stable and metastable adsorption sites for a single oxygen atom. (a) Two 2H-like surface-bridge sites labeled as 1-1-1 and 1-1-2, and three 3C-like surface-bridge sites labeled as 1-2-1, 1-2-2 and 1-2-3. (b) Two on-top sites labeled as 2-1 and 2-2. (c) Three subsurface-bridge sites labeled as 3-1, 3-2 and 3-3. The color code is the same as in Fig. \ref{fig_1120}. The highlighted atoms in (c) are the topmost atoms considered.}
	\label{fig_1O}
\end{figure}

In this subsection, we present and discuss the stable adsorption structures obtained for a single O-atom on the ($11\bar20$) surface. 
We start with the 15 initial configurations (Fig. \ref{fig_init-1O}), which cover all the possible adsorption sites for the one-O-atom adsorption.
Those structures include all the on-top sites and bridge sites for the topmost atoms along with the bridge sites between the topmost and subsurface atoms.
The surface structure of 4H-SiC ($11\bar20$) slab is relaxed before adding one oxygen atom 
and its atomic and electronic structures are in good agreement with previous work by Bertelli et al.\cite{SiC-surface} 
The initial structures along with the clean-surface structure, are schematically shown in Fig. \ref{fig_init-1O}.
After extensive geometry optimization, we have reached 10 stable and metastable adsorption structures (Fig. \ref{fig_1O}). 
They are classified into (1) surface-bridge site where O bridges two topmost surface atoms, 
(2) the on-top site where O sits on the topmost surface atom, and
(3) the subsurface-bridge site where O bridges the topmost surface atom and the subsurface atom.

In the surface-bridge sites, we have examined 7 possible initial configurations as in Figs. \ref{fig_init-1O} (b) and (c) 
and have found 5 stable and metastable structures shown in Fig. \ref{fig_1O} (a). 
As is clear from Fig. \ref{fig_1120db}, the stable surface-bridge sites are classified into two groups: 
In the first group the two surface dangling bonds are directed to the bridge site and in the second they show the different directions. 
We call the former and the latter as the 2H-like (labeled as 1-1-1 and 1-1-2) and the 3C-like (labeled as 1-2-1, 1-2-2 and 1-2-3), 
respectively since those configurations are the same as those on the corresponding surfaces of 2H-SiC and 3C-SiC. 
We have found two stable 2H-like bridge sites shown in Fig. \ref{fig_1O} (a): i.e., the on-bond site (labeled as 1-1-1) and the off-bond site (labeled as 1-1-2). 
In the on-bond site, the oxygen forms chemical bonding with C and Si atoms that are bonded with each other before oxygen adsorption. 
On the other hand, in the off-bond site, the adsorption takes place on the middle of C and Si atom that are not bonded before adsorption. 
In these two sites, the oxygen atoms are not located right above the C-Si bond but slightly dislodged. 
This is due to the directions of the two surface dangling bonds. As for the 3C-like bridge sites we have found the three stable sites shown in Fig. \ref{fig_1O}(a): 
i.e., the site between the k-site C and the k-site Si (labeled as 1-2-1), 
the site between the h-site C and the k-site Si (labeled as 1-2-2) and the site between the k-site C and the h-site Si (labeled as 1-2-3). 
In all the 3C-like bridge sites, the oxygen is located right above the C-Si bond. We have found that bridge sites between either Si atoms or C atoms are unstable.

There are four topmost atoms inequivalent to each other on 4H-SiC ($11\bar20$) surface. 
Then we have examined the four on-top sites as in Fig. \ref{fig_init-1O} (d). 
We have found that only two of the four are stable [Fig. \ref{fig_1O} (b)]: the site on the k-site Si (labeled as 2-1) and the site on the h-site C (labeled as 2-2).

Each of the topmost atom participates in the two bonds on the surface and the remaining bond with the subsurface atom. 
Then we have examined bridge sites between the topmost and the sub-surface atoms. 
We have found that the three of such four sites are stable as shown in Fig. \ref{fig_1O} (c): 
i.e., the subsurface-bridge sites with the topmost h-site Si (labeled as 3-1), k-site Si (labeled as 3-2) and h-site C (labeled as 3-3). 
In the subsurface-bridge sites of 4H-SiC ($11\bar20$), when the topmost atom is k-site, then subsurface atom is h-site and vice versa.

The relative stability among the adsorption structures obtained is assessed by the adsorption energy $E_{\text{ad}}$ which is defined as,
\begin{equation}
	E_{\text{ad}} = E_{s} + \frac{1}{2}E_{O_2} - E_{t}
\end{equation}
where $E_t$, $E_s$, and $E_{O_2}$ are the total energies of the O-adsorbed surface, the clean surface and the O$_2$ molecule, respectively. 
With this definition, the adsorption site with larger adsorption energy is more likely to be realized.

\begin{table}
	\caption{Adsorption energies for single O-atom at various stable sites. The superscripts for each site show the labels for the corresponding structures in Fig. \ref{fig_1O}. The NA in the table indicates that the structure have been found to be unstable, and it changed to nearby surface-bridge site after the geometry optimization.}
	\label{table_1O}
	\begin{ruledtabular}
	\begin{tabular}{lllll}
	 	\multicolumn{1}{c}{Sites} & & &\multicolumn{1}{c}{ID} & \multicolumn{1}{c}{$E_{\text{ad}}$} \\ \cline{1-5}
		\multicolumn{1}{c}{Surface-bridge site} & \multicolumn{1}{c}{2H-like} & \multicolumn{1}{c}{on-bond} & \multicolumn{1}{c}{1-1-1} &\multicolumn{1}{c}{2.35 eV} \\ \cline{3-5}
		 & & \multicolumn{1}{c}{off-bond} & \multicolumn{1}{c}{1-1-2} & \multicolumn{1}{c}{2.34 eV} \\ \cline{2-5}
		 & \multicolumn{1}{c}{3C-like} & \multicolumn{1}{c}{kC-kSi} &\multicolumn{1}{c}{1-2-1} & \multicolumn{1}{c}{1.74 eV} \\ \cline{3-5}
		 & & \multicolumn{1}{c}{hC-kSi} &\multicolumn{1}{c}{1-2-2} & \multicolumn{1}{c}{1.68 eV} \\ \cline{3-5}
		 & & \multicolumn{1}{c}{kC-hSi} &\multicolumn{1}{c}{1-2-3} & \multicolumn{1}{c}{1.57 eV} \\ \cline{1-5}
		\multicolumn{1}{c}{On-top site} & \multicolumn{1}{c}{Si} & \multicolumn{1}{c}{h-site} & \multicolumn{1}{c}{NA} & \multicolumn{1}{c}{-} \\ \cline{3-5}
		 &  & \multicolumn{1}{c}{k-site} & \multicolumn{1}{c}{2-1} & \multicolumn{1}{c}{1.04 eV} \\ \cline{2-5}
		 & \multicolumn{1}{c}{C} & \multicolumn{1}{c}{h-site} & \multicolumn{1}{c}{2-2} & \multicolumn{1}{c}{1.56 eV} \\ \cline{3-5}
		 &  & \multicolumn{1}{c}{k-site}& \multicolumn{1}{c}{NA} & \multicolumn{1}{c}{-} \\ \cline{1-5}
		\multicolumn{1}{c}{Subsurface-bridge site} & \multicolumn{1}{c}{Si} & \multicolumn{1}{c}{h-site} & \multicolumn{1}{c}{3-1} & \multicolumn{1}{c}{0.90 eV} \\ \cline{3-5}
		 &  & \multicolumn{1}{c}{k-site} &\multicolumn{1}{c}{3-2} & \multicolumn{1}{c}{1.21 eV} \\ \cline{2-5}
		 & \multicolumn{1}{c}{C} & \multicolumn{1}{c}{h-site} & \multicolumn{1}{c}{3-3} & \multicolumn{1}{c}{1.27 eV} \\ \cline{3-5}
		 &  & \multicolumn{1}{c}{k-site} & \multicolumn{1}{c}{NA} & \multicolumn{1}{c}{-} \\
	\end{tabular}
	\end{ruledtabular}
\end{table}

The calculated adsorption energies are shown in Table \ref{table_1O}. 
We have found that the calculated adsorption energy is in the range of 0.90 to 2.35 eV, indicating that the O-atom adsorption is the exothermic reaction. 
The most stable adsorption structure is achieved at the surface-bridge site. 
This is due to the efficient annihilation of the dangling bonds at the topmost surface atoms: 
At the bridge site, the O is capable of terminating two surface dangling bonds. 
The adsorption energies at the 2H-like bridge sites are larger than those at the 3C-like bridge sites by more than a half eV. 
This is due to the strains caused by the bond-angle deviation at the 3C-like sites. 
For on-top sites, only the k-site Si and the h-site C are found to be stable structures. 
The O-atom adsorption on the C on-top site is more favorable than the Si on-top site. 
This is also true for the subsurface-bridge site, indicating that the termination of C dangling bond is energetically more favorable. 
There are three sites labeled as NA in Table \ref{table_1O}. 
This indicates that the final structure after the geometry optimization is found to be the nearby surface-bridge sites. 
We have now unequivocally revealed the energy landscape for the O-atom adsorption on the ($11\bar20$) surface of 4H-SiC.

\subsection{Two-oxygen-atom adsorption}

\begin{figure}
	\includegraphics[width=0.5\textwidth]{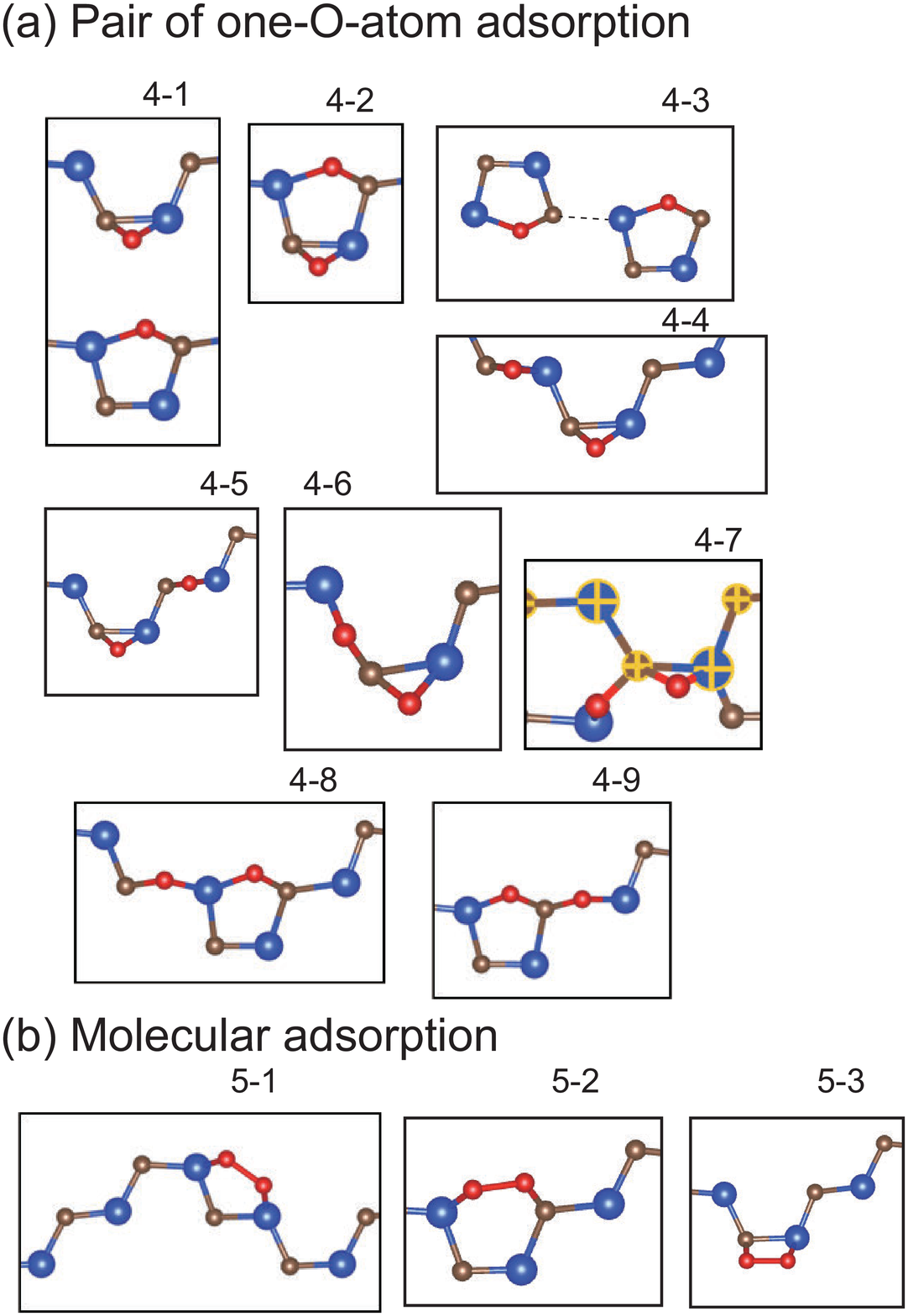}
	\caption{Top views of stable adsorption structures for two oxygen atoms. (a) Pair-structures consisting of the stable 2H-like adsorption site and other plausible site: The 4-1 and 4-2 = (1-1-1) + (1-1-2), the 4-3 = (1-1-2) + (1-1-2), the 4-4 and 4-5 = (1-1-1) + (1-2-1), the 4-6 = (1-1-1) + (1-2-3) and the 4-7 = (1-1-1) + (3-3). The 4-8 and 4-9, consisting of (1-1-2) and (1-2-1), are unusual structures (see text) which shows relatively large adsorption energies. (b) Molecular adsorption where the O-O bond survives after the adsorption. The color code is the same as in Fig. \ref{fig_1120}}.
	\label{fig_2O}
\end{figure}

In this subsection, we present the stable adsorption structures for two O-atom near to each other on the ($11\bar20$) surface to mimic the further oxidation at the initial stage. 
The plausible structure with large adsorption energy may be a pair of the most stable adsorption structures for the single O-atom. 
Then we consider the 2H-like surface-bridge site, 1-1-1 or 1-1-2, as a partner of such pair and seek for the plausible site for the next oxygen. 
After the extensive geometry optimization, we have found 9 stable pairs shown in Fig. \ref{fig_2O} (a) as labeled from 4-1 to 4-9. 
A measure for the relative stability among the stable structures is the adsorption energy defined as
\begin{equation}
	E_{\text{ad}} = E_{s} + E_{O_2} - E_{t}.
\end{equation}
The calculated adsorption energies for the 9 stable geometries are tabulated in Table \ref{table_2O}.

We have found that the adsorption energy $E_{\text{ad}}$ for the two-O-atom is generally smaller than the sum of the corresponding adsorption energies $E_{\text{sum}}$ for the single O-atom. 
This is due to the strain energies caused by the nearby O atoms. 
For instance, the structures having the largest sum of the adsorption energies for the single O-atom are the pairs of 2H-like surface-bridge sites labeled 4-1 and 4-2. 
The calculated values for these pairs are smaller than the corresponding sum by 0.20 and 0.47 eV, respectively, due to the competing structural relaxation. 
Interesting exceptions are found in the structures labeled 4-8 and 4-9: 
i.e., the pair of the 2H-like off-bond site 1-1-2 and the 3C-like surface-bridge site 1-2-1 of which the sum of the adsorption energies is 4.08 eV. 
Our structural optimization for these pairs leads to the structures (Fig. \ref{fig_2O}) in which a characteristic Si-O-C-O-Si network is formed. 
The calculated adsorption energy for the 4-8 is larger than the corresponding sum by 0.98 eV, 
and even larger than the pair 4-2 of the 2H-like surface-bridge sites by 0.57 eV (Table \ref{table_2O}).

We have also examined several molecular adsorptions that have no counterparts in the single O-atom adsorption. 
The calculated adsorption energies for the molecular adsorption are generally small, indicating that the O$_2$ molecule is likely to be dissociated on the surface. 
However, dynamical aspects should be examined to reach a definite conclusion.

\begin{table}
	\caption{Adsorption energies $E_{\text{ad}}$ for two-O-atom at various sites. $E_{\text{sum}}$ is the sum of the adsorption energies for the respective O atoms forming the pair (see text).}
	\label{table_2O}
	\begin{ruledtabular}
	\begin{tabular}{llll}
		\multicolumn{1}{c}{ID} & \multicolumn{1}{c}{Pair} & \multicolumn{1}{c}{$E_{\text{sum}}$} & \multicolumn{1}{c}{$E_{\text{ad}}$} \\ \cline{1-4}
		\multicolumn{1}{c}{4-1} & \multicolumn{1}{c}{1-1-1 and 1-1-2} & \multicolumn{1}{c}{4.69 eV} & \multicolumn{1}{c}{4.22 eV} \\
		\multicolumn{1}{c}{4-2} & & & \multicolumn{1}{c}{4.49 eV} \\
		\multicolumn{1}{c}{4-3} & \multicolumn{1}{c}{1-1-1 and 1-1-1} & \multicolumn{1}{c}{4.68 eV} & \multicolumn{1}{c}{4.27 eV} \\
		\multicolumn{1}{c}{4-4} & \multicolumn{1}{c}{1-1-1 and 1-2-1} & \multicolumn{1}{c}{4.09 eV} & \multicolumn{1}{c}{3.96 eV} \\ 
		\multicolumn{1}{c}{4-5} &  &  & \multicolumn{1}{c}{3.86 eV} \\
		\multicolumn{1}{c}{4-6} & \multicolumn{1}{c}{1-1-1 and 1-2-3} & \multicolumn{1}{c}{3.91 eV} & \multicolumn{1}{c}{3.99 eV} \\ 
		\multicolumn{1}{c}{4-7} & \multicolumn{1}{c}{1-1-1 and 3-3} & \multicolumn{1}{c}{3.62 eV} & \multicolumn{1}{c}{3.95 eV} \\
		\multicolumn{1}{c}{4-8} & \multicolumn{1}{c}{1-1-2 and 1-2-1} & \multicolumn{1}{c}{4.08 eV} & \multicolumn{1}{c}{5.06 eV} \\
		\multicolumn{1}{c}{4-9} & &  & \multicolumn{1}{c}{4.66 eV} \\ \cline{1-4}
		\multicolumn{1}{c}{5-1} & \multicolumn{1}{c}{-} & \multicolumn{1}{c}{-} & \multicolumn{1}{c}{1.93 eV} \\
		\multicolumn{1}{c}{5-2} & & & \multicolumn{1}{c}{2.83 eV} \\
		\multicolumn{1}{c}{5-3} & & & \multicolumn{1}{c}{2.44 eV} \\
	\end{tabular}
	\end{ruledtabular}
\end{table}

\section{Result II: adsorption and Dissociation of O$_2$ molecule}

So far, We have unequivocally identified the adsorption structures in terms of their energetics for the single and 
two-O-atom on the basis of their relative total energies. 
Since these oxidation processes are highly dynamical, temperature and entropy effects should be included to get a complete insight into the oxidation phenomenon.
This is the scope of the CPMD simulations done to investigate the adsorption and then dissociation of the O$_2$ molecule on the 4H-SiC ($11\bar20$) surface. 
Although the ground state of an O$_2$ molecule in the gas phase is triplet, 
we consider the approaching molecule in a spin singlet state, implying that for oxidation, the starting position of 
O$_2$ is sufficiently close to the surface so that the spin polarization has already disappeared.

\subsection{O$_2$ molecule adsorption}

\begin{figure*}
	\includegraphics[width=1\textwidth]{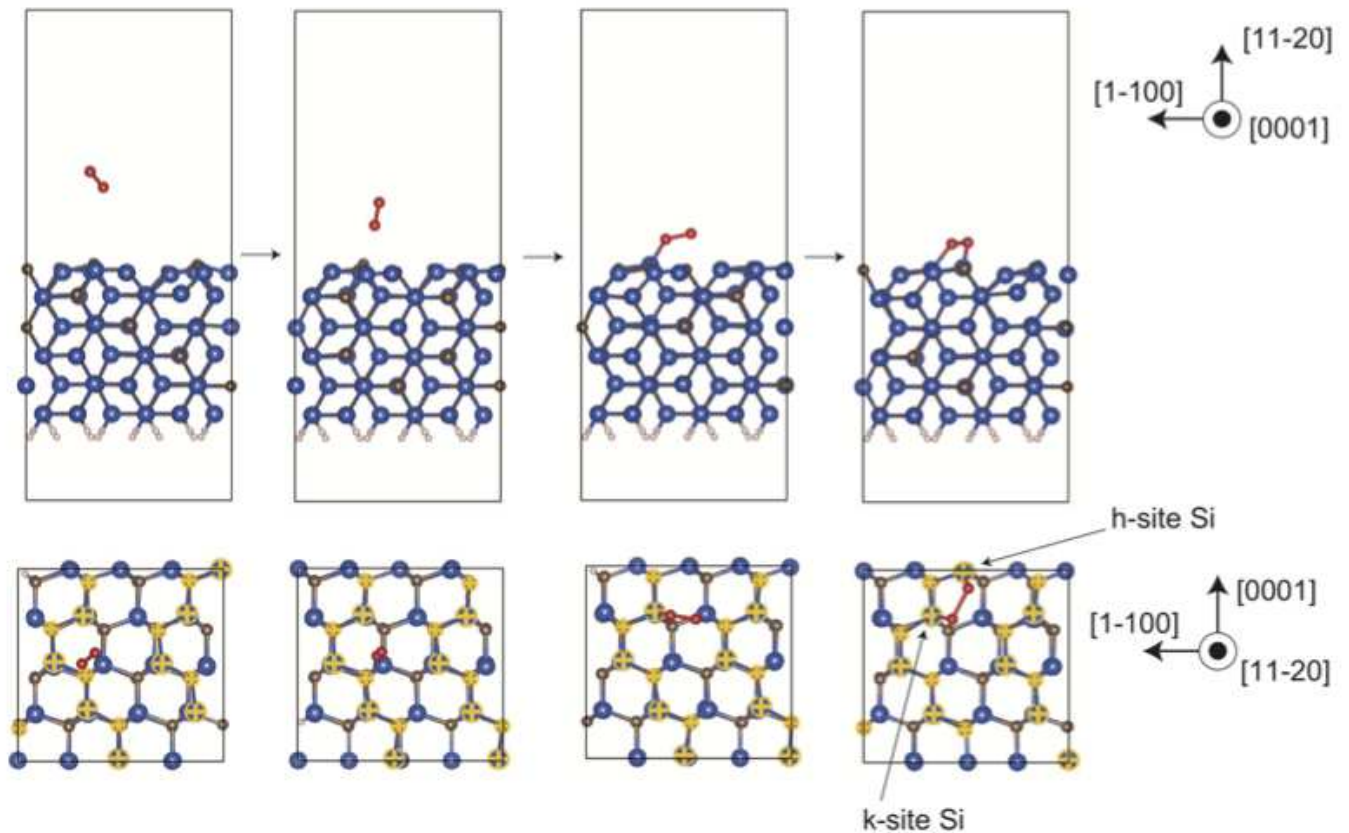}
	\caption{Adsorption of an O$_2$ molecule to ($11\bar20$) surface represented by snapshots during the CPMD simulations. Side views (upper panels) and top views (lower panels). When the velocity of O$_2$ molecule toward the surface is sufficiently small, the molecule is adsorbed on the surface forming Si-O-O-Si structure. The color code is the same as in Fig. \ref{fig_1120}}
	\label{fig_Si-O-O-Si}
\end{figure*}

In our dynamical simulations, we first equilibrate the system at room temperature (300 K). 
Then we introduce a slow oxygen molecule with a kinetic energy of 0.1 meV in the vicinity of the surface. 
With this relatively small kinetic energy, the oxygen molecule approaches the energetically most 
favorable site among the accessible geometrical configurations, irrespective of the initial direction of the velocity. 
In this case, we observed that the O$_2$ molecule first attacks the k-site Si and then the nearby h-site Si to 
form a Si-O-O-Si bridge structure (Fig. \ref{fig_Si-O-O-Si}). As this process occurs spontaneously without 
imposing any constraints to drive a specific reaction, the energy barrier for the adsorption of the single O$_2$ 
molecule to 4H-SiC ($11\bar20$) should have as an upper bound value of 25.7 meV, i.e. the system temperature at 300K. 
In the Si-O-O-Si bridge structure, the O-O distance is stretched beyond the bond length of the O$_2$ molecule. 
However, our analysis of the electron density indicates clearly that the chemical bond is preserved. Therefore the 
molecular adsorption occurs on the ($11\bar20$) surface at the initial stage of the process and this adsorption 
does not result in a molecular dissociation. The structure realized is the same as 5-1 in Fig. \ref{fig_2O}. 
Moreover, these dynamical simulations indicate a higher affinity of Si atoms than C atoms to the O.

\subsection{O$_2$ dissociation}

\begin{figure}
	\includegraphics[width=0.5\textwidth]{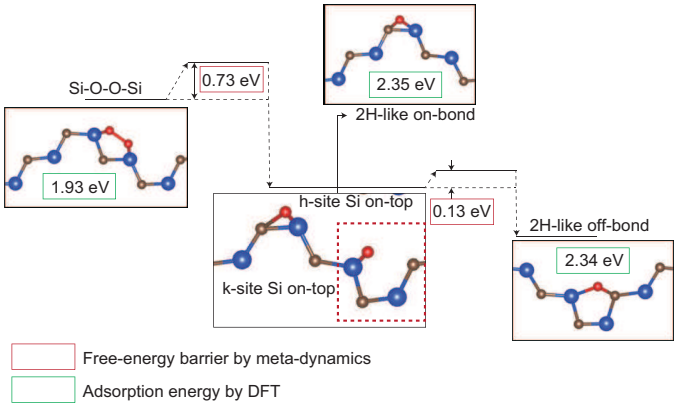}
	\caption{Structural transformation during O$_2$ dissociation at ($11\bar20$) surface. Starting from O$_2$ molecular adsorption structure (left panel), the O-O bond breaks with 0.73 eV free energy barrier (middle bottom panel). The h-site Si on-top site is not stable for O atom so that it is transformed into 2H-like on-bond structure (middle top panel). It takes 0.13 eV for the k-site Si on-top structure to go to more stable 2H-like off-bond structure (right panel).}
	\label{fig_dissociation}
\end{figure}

In the previous subsection, we have clarified that the molecular O$_2$ adsorption takes place on the ($11\bar20$) 
surface at room temperature. On the other hand, our DFT calculations in section III show that the adsorption energy 
for an O$_2$ molecule is much smaller than the dissociative adsorption of two-O-atom (Table \ref{table_2O}). 
Then the dissociation of O$_2$ is expected to occur and this process is indispensable in order for the oxidation to proceed. 
This is very different from the adsorption of O$_2$ molecule on 4H-SiC ($000\bar1$) surface where the dissociation occurs 
spontaneously.\cite{MO-unpublished} We have adopted the final structure obtained from the CPMD simulation for the O$_2$ molecular adsorption 
in the preceding subsection as an initial structure, and then conducted 
further CPMD simulations complemented with the meta-dynamics approach for the sampling of the reaction pathways and 
to work out corresponding free-energy barriers for the dissociation. 
The selected CVs were the O-O distance and the distance between an O atom and a C atom at the k-site. We have found that 
the O$_2$ molecule is dissociated and the resulting O atoms are adsorbed at the on-top sites of the h-site Si and the 
k-site Si, respectively. The activation barrier for this O$_2$ dissociation is 0.73 eV (Fig. \ref{fig_dissociation}). 
From the DFT calculations for the single O atom adsorption in the previous section, the O atom at the on-top site of the 
h-site Si is unstable (Table \ref{table_1O}). Considering that the 2H-like on-bond bridge structure 1-1-1 is reached by 
small dislodgment from the on-top site on the h-site Si [Fig. \ref{fig_1O} (a)] and the 1-1-1 structure is the most stable 
adsorption site, the O atom at the on-top site on the h-site Si after the O$_2$ dissociation transforms 
to the nearby bridge structure 1-1-1 with no appreciable energy barrier. As for the O atom at the on-top site on the k-site Si which is metastable, we have 
performed CPMD simulations and found that it transforms to another 2H-like off-bond bridge site 1-1-2 with the activation 
energy of 0.13 eV (Fig. \ref{fig_dissociation}).

Dynamical simulations presented in this section, along with the static calculations in the preceding section III, 
clarify that oxygen is adsorbed in an undissociated molecular form at the beginning of the process and then dissociated 
toward more stable atomic-oxygen adsorption structures upon the overcoming of a free-energy barrier of 1 eV or less. 
The results also reflect the higher oxygen affinity of Si than C atoms. 
We have found several stable structures for the pair of the resultant O atoms. 
This multi-stability is the characteristics at the initial stage of the oxidation on ($11\bar20$) surface.

\section{Result III: Desorption of CO molecule}

The remaining important process in the initial stage of the oxidation of SiC is the removal of carbon atoms by 
specific reactions, which is a necessary step for SiO$_2$ films to be formed. By examining the stable structures carrying 
a single O-atom and two-O-atom discussed in the previous sections [Figs. \ref{fig_1O} and \ref{fig_2O}], we consider 
the desorption of a CO molecule from the surface as one of the important reaction process for the C removal. 
To this aim, we resort to CPMD simulation to inspect the desorption of CO and we compare this CO desorption on the ($11\bar20$) surface 
with the analogous process on the ($000\bar1$) surface to illustrate the peculiarities of the ($11\bar20$) surface.

We examine the CO desorption on the oxygen adsorbed SiC surfaces with and without hydrogen termination. 
This is motivated by the interest in the role of the surface dangling bonds in the desorption process on one hand. 
On the other, the H-terminated surface mimics the SiO$_2$/SiC interface where the Si-O-C network is likely to exist 
during the oxidation with the dangling bonds being not near to the network.

\subsection{CO desorption from the clean surface}

\begin{figure}
	\includegraphics[width=0.5\textwidth]{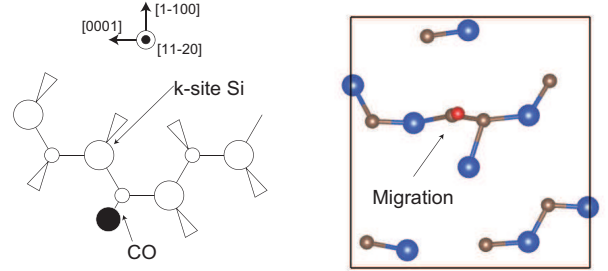}
	\caption{(Left panel) Schematic view of the configuration of the dangling bonds. (Right panel) Top view of the structure after the migration of CO at the h-site on ($11\bar20$) surface.}
	\label{fig_after-migration}
\end{figure}

Among the stable and metastable oxygen adsorbed structures, the structure labeled 2-2 in Fig. \ref{fig_1O} is the most probable candidate structure for the CO desorption. 
As stated in section II, we have chosen the coordination number of the target C atom as the collective variable, 
i.e., the reaction coordinate, in the CPMD simulations with the meta-dynamics. 
We have found that the desorption process on the ($000\bar1$) is just the elongation of the C-Si distance, 
leading to the CO desorption. 
The calculated free-energy barriers are 1.51 eV and 1.68 eV for the h-site and k-site carbon respectively. 
As for the ($11\bar20$) surface, we have found that the desorption of CO takes place for the k-site C. 
The calculated free-energy barrier is 2.54 eV, which is much higher than that on the ($000\bar1$) surface (Table \ref{table_clean}). 
As for the h-site, we have found that the free-energy barrier for the surface migration of the CO is much lower than the desorption.
We have found that the CO unit at the h-site migrates to the bridge site shown in Fig. \ref{fig_after-migration}. 
The obtained free-energy barrier is 0.59 eV. This relatively low energy barrier is caused by the surface dangling bonds at the k-site Si (Fig. \ref{fig_after-migration}). 
It is noteworthy that this migration process on the ($11\bar20$) surface corresponds to the outward migration of the CO unit for the ($000\bar1$) surface of SiC, 
which may assists  in the carbon annihilation from the interface.

\begin{table}
	\caption{Desorption energy barriers for CO on cleavage surface.}
	\label{table_clean}
	\begin{ruledtabular}
	\begin{tabular}{lll}
		\multicolumn{1}{c}{Surface orientation} & \multicolumn{1}{c}{($11\bar20$)} & \multicolumn{1}{c}{($000\bar1$)} \\ \cline{1-3}
		\multicolumn{1}{c}{h-site} & \multicolumn{1}{c}{0.59 eV (Migration)} & \multicolumn{1}{c}{1.51 eV} \\
		\multicolumn{1}{c}{k-site} & \multicolumn{1}{c}{2.54 eV} & \multicolumn{1}{c}{1.68 eV} \\
	\end{tabular}
	\end{ruledtabular}
\end{table}

\subsection{CO Desorption from the H-terminated surface}

We next consider the CO desorption from the hydrogen terminated oxidized surface. 
The energy barriers obtained by our meta-dynamics-enhanced CPMD simulations are shown in Table \ref{table_H-term}. 
Specifically, the free-energy barriers for the desorption from the ($11\bar20$) surface are 2.61 and 2.41 eV for h-site 
and k-site respectively. For comparison, on the ($000\bar1$) surface the calculated values are 2.62 and 2.59 eV, respectively. 
The free-energy barriers for the CO desorption from a H-terminated surface are rather similar on both surfaces and site type, either h- or k-sites. 
It is noteworthy that the obtained values are also close to the desorption barriers on the H-free ($11\bar20$) surface (Table \ref{table_clean}). 
It is also remarkable that the desorption barriers from the H-free ($000\bar1$) surface are definitely smaller, about one and a half eV. 
We have found that this trend in the desorption barriers is parallel to the trend in the total energies of the final geometries.

By analyzing the electronic structures, we have found that this trend is caused by the presence or absence of the partially filled C dangling bonds. 
For the desorption from the H-free ($000\bar1$) surface, three Si dangling bonds are generated after the CO desorption. 
Each electron in the Si dangling bonds are transferred to the dangling bonds of the topmost C atoms which are located in energy below the Si dangling bond 
(C has larger electron affinity than Si) and partially filled, leading to the surface states. 
This electronic energy gain is the reason for the lower final-state and transition-state energy. For the ($11\bar20$) surface, there exist an equal amount of both Si and C dangling bonds. 
Each electron of the Si dangling bond is already transferred to the C dangling bond, leading to the gap of the surface states, before the CO desorption. 
Then it is obvious that the electronic energy gain is absent. For the H-terminated surface, 
this mechanism of the electron acceptance by the C dangling bonds and the resultant energy gain is invalid. 
We just observe the structural relaxation of the two of the three topmost Si atoms after the CO desorption on the ($000\bar1$) surface, 
which is minor in the energetics.

\begin{table}
	\caption{Desorption energy barriers for CO on H-terminated surface.}
	\label{table_H-term}
	\begin{ruledtabular}
	\begin{tabular}{lll}
		\multicolumn{1}{c}{Surface orientation} & \multicolumn{1}{c}{($11\bar20$)} & \multicolumn{1}{c}{($000\bar1$)} \\ \cline{1-3}
		\multicolumn{1}{c}{h-site} & \multicolumn{1}{c}{2.61 eV} & \multicolumn{1}{c}{2.62 eV} \\
		\multicolumn{1}{c}{k-site} & \multicolumn{1}{c}{2.41 eV} & \multicolumn{1}{c}{2.59 eV} \\
	\end{tabular}
	\end{ruledtabular}
\end{table}

\section{Conclusion}

We have performed static and dynamic calculations based on DFT to reveal microscopic mechanisms of oxidation of SiC ($11\bar20$) surface at its initial stage.

Firstly, we have identified stable and metastable adsorption structures for a single oxygen atom and two oxygen atoms. 
This was done by exhaustive search for possible geometries by the static DFT calculations. 
We have found 10 stable structures for the single oxygen atom and 12 stable structures for the two oxygen atoms. 
From the calculated adsorption energies, we have found that the two particular surface-bridge sites where the local configurations are wurtzite-like (2H-like) 
are the most stable adsorption sites for the single oxygen atom. 
Other surface-bridge sites where the local structures are zincblende-like (3C-like) are found to be the next stable adsorption sites. 
We have also found that the on-top sites and the subsurface-bridge sites show relatively small adsorption energies. 
The energetics obtained shows that oxygen tends to attack the carbon, since the termination of carbon dangling bonds is energetically more favorable.
For the two-oxygen-atom adsorption, we have found that the pairs of the surface-bridge sites are the most stable adsorption sites. 
Yet the adsorption energies are generally smaller than the sum of the adsorption energies at the surface-bridge sites for the single O atom 
due to the strains caused by the insertion of O atoms. 
Interesting exception is the pair of the 2H-like off-bond surface-bridge site and the 3C-like surface-bridge site. 
This configurations are composed of the -C-O-Si-O-C or -Si-O-C-O-Si- network along [0001] direction, 
thus releasing the strain energy. Our results are indicative of the importance of the surface-bridge sites in the initial stage of the oxidation process. 
In addition to that, our finding of multi-stability of the adsorption structures is the characteristics of the high-Miller index surface 
like the ($11\bar20$) surface where various sites with different symmetries provide comparable accommodation sites for the oxygen.

Secondly, we have performed CPMD simulations to inspect reaction pathways 
and to compute the corresponding free-energy barriers for the adsorption and dissociation of the oxygen molecule and also the desorption of the CO molecule from the ($11\bar20$) surface. 
The free-energy barriers have been obtained by the CPMD simulations combined with the meta-dynamics. 
We have found that the O$_2$ molecule is adsorbed on the surface-bridge site where each of the O atom forms a bond with the topmost Si atom. 
This shows higher affinity of Si than C with O dynamically. Our calculation also shows that the adsorption energy of the O$_2$ molecule is much smaller that that of O atoms. 
Our CPMD calculations have indeed revealed that the adsorbed O$_2$ molecule is dissociated with the free-energy barrier of 0.7 eV 
and the resultant two O atoms are eventually adsorbed at the 2H-like surface-bridge sites.

The desorption of the CO molecule is the fundamental process in the annihilation of C during the oxidation. 
We have performed the CPMD simulations for the CO desorption from the ($11\bar20$) surface and also from the ($000\bar1$) surface for comparison. 
We have found that C is removed in the form of CO molecule.
The calculated free-energy barriers for the desorption are 2.4$\sim$2.6 eV for most cases, 
and this is attributed to be the free-energy needed to break three Si-C bonds.
We have examined both the clean and the hydrogen-terminated oxidized surface to clarify the role of the surface dangling bonds in the CO desorption. 
We have found that the dangling bonds at the surface play an important role during the oxidation; they can reduce the activation barrier for certain reactions and can also change the dominant reaction pathways.
For the ($000\bar1$) surface, an interesting reduction of the desorption barrier by about 1 eV is observed.
Our analyses have clarified that this reduction is caused by the electron transfer from the Si dangling bonds to the C dangling bonds upon the CO desorption.
For the ($11\bar20$) surface, a change from desorption of CO to surface migration occurs when dangling bonds are positioned near the CO.
Our findings about the dangling-bond effects are crucial when constructing further oxidized theoretical SiO$_2$/SiC interface model 
to study the chemical reactions at the interface.
The dangling bonds at the interface need to be treated carefully because they can enhance certain reactions.

\begin{acknowledgements}
This work was partly supported by the project for Priority Issue "Creation of new functional devices and high-performance materials to support next-generation industries)"
to be tackled by using Post-K Computer, conducted by Ministry of Education, Culture, Sports, Science and Technology, Japan, 
and by the JSPS KAKENHI Grant Number 15H03601. 
Computations were performed at the Supercomputer Center at the Institute for Solid State Physics, The University of Tokyo, 
The Research Center for Computational Science, National Institutes of Natural Sciences, the Center for Computational Science, 
University of Tsukuba, and at the HPC Mesocenter (Equip@Meso) of the University of Strasbourg, France. 
H. L. was supported by Japan Society for the Promotion of Science through Program for Leading Graduate Schools (MERIT). 
M. B. acknowledges LaBex "Nanoparticles Interacting with their Environment" ANR-11-LABX-0058\_NIE and GENCI under allocation x2016096092.
\end{acknowledgements}

\bibliography{ref}

\end{document}